\documentstyle[12pt]{article}
\begin{document}

\renewcommand{\thefootnote}{\fnsymbol{footnote}}
\newcommand{\beq}{\begin{equation}}
\newcommand{\eeq}{\smallskip\end{equation}}
\newcommand{\0}{\mbox{$|0\rangle$}}
\newcommand{\1}{\mbox{$|1\rangle$}}
\newcommand{\2}{\otimes}
\newcommand{\zz}{\mbox{$|z\rangle$}}
\newcommand{\ao}{\mbox{$|a=0\rangle$}}
\newcommand{\ket}{\rangle}
\newcommand{\bra}{\langle}
\newcommand{\one}{\leavevmode\hbox{\small1\kern-3.8pt\normalsize1}}

\vspace*{16mm}
\begin{center}{\large{\bf Unitary Dynamics for Quantum
Codewords}}\\[8mm] 
Asher Peres\footnote{Permanent address: Technion---Israel Institute of
Technology, 32\,000 Haifa, Israel}\\[4mm] {\sl
Institute for Theoretical Physics\\ University of California\\ Santa
Barbara, CA 93106}\end{center}\medskip

\begin{quote}{\small A quantum codeword is a redundant representation
of a logical qubit by means of several physical qubits. It is
constructed in such a way that if one of the physical qubits is
perturbed, for example if it gets entangled with an unknown environment,
there still is enough information encoded in the other physical qubits
to restore the logical qubit, and disentangle it from the environment.
The recovery procedure may consist of the detection of an error
syndrome, followed by the correction of the error, as in the classical
case. However, it can also be performed by means of unitary operations,
without having to know the error syndrome.

Since quantum codewords span only a restricted subspace of the complete
physical Hilbert space, the unitary operations that generate quantum
dynamics (that is, the computational process) are subject to
considerable arbitrariness, similar to the gauge freedom in quantum
field theory. Quantum codewords can thus serve as a toy model for
investigating the quantization of constrained dynamical
systems.}\end{quote}

\bigskip\noindent{\bf 1. Introduction}\bigskip

\noindent In {\it classical\/} communication and computing systems,
logical bits, having values 0 or 1, are implemented in a highly
redundant way by bistable elements, such as magnetic domains. The
bistability is enforced by coupling the bit carriers to a dissipative
environment. Errors may then occur, because of thermal fluctuations
and other hardware imperfections. To take care of these errors, various
correction methods have been developed [1], involving the use of
redundant bits (that are implemented by additional bistable elements).

In {\it quantum\/} communication and computing, the situation is more
complicated: in spite of their name, the logical ``qubits'' (quantum
binary digits) are not restricted to the discrete values 0 and 1. Their
value can be represented by any point on the surface of a Poincar\'e
sphere. Moreover, any set of qubits can be in an {\it entangled\/}
state: none of the individual qubits has a pure quantum state, it is
only the state of all the qubits together that is pure [2].

The qubits of a quantum computer are materialized by single
quanta, such as trapped ions~[3]. Their coupling to a dissipative
environment (which was the standard stabilizing mechanism for classical
bits) is now to be avoided as much as possible, because it readily
leads to decoherence, namely to the loss of phase relationships. Yet,
disturbances due to the environment cannot be completely eliminated:
e.g., even if there are no residual gas molecules in the vacuum of an
ion trap, there still are the vacuum fluctuations of the quantized
electromagnetic field, which induce spontaneous transitions between the
energy levels of the ions. Therefore, error control is an essential
part of any quantum communication or computing system.

\renewcommand{\thefootnote}{\arabic{footnote}}
\setcounter{footnote}{0}

This goal is much more difficult to achieve than classical error
correction, because qubits cannot be read, or copied, or duplicated,
without altering their quantum state in an unpredictable way [4]. The
feasibility of quantum error correction, which for some time had been
in doubt, was first demonstrated by Shor [5]. As in the classical case,
{\it redundancy\/} is an essential element, but this cannot be a simple
repetitive redundancy, where each bit has several identical replicas
and a majority vote is taken to establish the truth. This is because
qubits, contrary to ordinary classical bits, can be {\it entangled\/},
and usually they are. As a trivial example, in the singlet state of two
spin-$1\over2$ particles, each particle, taken separately, is in a
completely random state. Therefore, comparing the states of
spin-$1\over2$ particles that belong to different (redundant) singlets
would give no information whatsoever.

All quantum error correction methods [5--9] use several physical qubits
for representing one logical qubit. These physical qubits are prepared
in a carefully chosen, highly entangled state. None of these qubits,
taken alone, carries any information. However, a large enough subset of
them may contain a sufficient amount of information, encoded in
relative phases, for determining and exactly restoring the state of the
logical qubit, including its entanglement with the other logical qubits
in the quantum computer.

In this article, I review the quantum mechanical principles
that make error correction possible. (I shall not discuss how to actually
design new codewords; the most efficient techniques involve a
combination of classical coding theory and of the theory of finite
groups.) Since quantum codewords span only a restricted subspace of the
complete physical Hilbert space, the unitary operations that generate
the quantum dynamical evolution (that is, the computational process)
are subject to considerable arbitrariness. The latter is similar to the
gauge freedom in quantum field theory. Quantum codewords can thus serve
as a simple toy model for investigating the quantization of constrained
dynamical systems, such as field theories with gauge groups.\\[7mm]

\noindent{\bf 2. Encoding and decoding}\bigskip

\noindent In the following, I shall usually consider codewords that
represent a single logical qubit. It is also possible, and perhaps it
may be more efficient, to encode several qubits into larger codewords.
However, no new physical principles are involved in this, and the
simple case of a single qubit is sufficient for illustrating these
principles.

The quantum state of a single logical qubit will be denoted as

\beq \psi=\alpha\,\0+\beta\,\1, \label{qubit}\eeq
where the coefficients $\alpha$ and $\beta$ are complex numbers. The
symbols \0\ and \1\ represent any two orthogonal quantum states, such as
``up'' and ``down''  for a spin, or the ground state and an excited
state of a trapped ion.

In a quantum computer, there are many logical qubits, typically in a
collective, highly entangled state, and any particular qubit has no
definite state. I shall still use the same symbol $\psi$ for
representing the state of the entire computer, and Eq.~(\ref{qubit})
could now be written as

\beq \psi=|\alpha\ket\otimes\0+|\beta\ket\otimes\1, \label{computer}\eeq
where one particular qubit has been singled out for the discussion, and
the symbols $|\alpha\ket$ and $|\beta\ket$ represent the collective
states of all the other qubits, that are correlated with \0\ and \1,
respectively.  However, to simplify the notation and improve
readability, I shall still write the computer state as in
Eq.~(\ref{qubit}). In the following, Dirac's ket notation will in
general {\it not\/} be used for generic state vectors (such as $\psi$,
$\alpha$, $\beta$) and the $\otimes$ sign will sometimes be omitted,
when the meaning is clear. Kets will be used only for denoting basis
vectors such as \0\ and \1, and their direct products.  The latter will
be labelled by binary numbers, such as

\beq |9\ket\equiv|01001\ket\equiv\0\otimes\1\otimes\0\otimes\0\otimes\1. 
\eeq

In order to encode the qubit $\psi$ in Eq.~(\ref{qubit}), we intoduce an
auxiliary system, called {\it ancilla\/},\footnote{This is the Latin
word for housemaid.} initially in a state $|000\ldots\,\ket$. The
ancilla is made of $n$ qubits, and we can use the mutually orthogonal
vectors $|a\ket$, with $a=0$, 1, \ldots\,, $2^n-1$ (the number $a$ being
written in binary notation) as a basis for its quantum states. These
labels are called {\it syndromes\/}, because, as we shall see, the
presence of an ancilla state with $a\neq0$ may serve to identify an
error in the encoded state that represents $\psi$.

Encoding is a unitary transformation, $E$, performed on a physical qubit
and its ancilla together:

\beq \zz\otimes\ao\to E\,\Bigl(\zz\otimes\ao\Bigr)\equiv|Z_0\ket,\eeq
where \zz\ means either \0\ or \1. This unitary transformation is
executed by a quantum circuit (an array of quantum gates). However, from
the theorist's point of view, it is also convenient to consider
$\zz\otimes\ao$ and $|Z_0\ket$ as two different representations of the
same qubit \zz: its logical representation, and its physical
representation. The first one is convenient for discussing matters of
principle, such as quantum algorithms, while the physical
representation is the one where qubits are actually materialized by
distinct physical systems (and the latter are the ones that may be
subject to independent errors).\footnote{These two different
representations are analogous to the use of normal modes vs.\ local
coordinates for describing the small oscillations of a mechanical
system [10]. One description is mathematically simple, the other one is
related to directly accessible quantities.}\\[7mm]

\noindent{\bf 3. Error correction}\bigskip

\noindent If there are $2^n$ syndromes (including the null syndrome for
no error), it is possible to identify and correct up to $2^n-1$
different errors that affect the physical qubits, with the help of a
suitable decoding method, as explained below. Let $|Z_a\ket$, with
$a=0$, \ldots\,, $2^n-1$, be a complete set of orthonormal vectors
describing the physical qubits of which the codewords are made:
$|0_0\ket$ and $|1_0\ket$ are the two error free states that represent
\0\ and \1, and all the other $|0_a\ket$ and $|1_a\ket$ are the results
of errors (affecting one physical qubit in the codeword, or several
ones, this does not matter at this stage).  These $|Z_a\ket$ are
defined in such a way that $|0_a\ket$ and $|1_a\ket$ result from the
{\it same\/} errors in the physical qubits of $|0_0\ket$ and $|1_0\ket$
(for example, the third qubit is flipped). We thus have two complete
orthonormal bases, $\zz\otimes|a\ket$ and $|Z_a\ket$. These two bases
uniquely define a unitary transformation $E$, such that

\beq E\,(\zz\otimes|a\ket)=|Z_a\ket, \eeq
and

\beq E^\dagger\,|Z_a\ket=\zz\otimes|a\ket, \label{decod} \eeq
where $a$ runs from 0 to $2^n-1$. Thus, $E$ is the encoding matrix, and
$E^\dagger$ is the decoding matrix. If the original and corrupted
codewords are chosen in such a way that $E$ is a real orthogonal matrix
(not a complex unitary one), then $E^\dagger$ is the transposed matrix,
and therefore $E$ and $E^\dagger$ are implemented by the {\it same\/}
quantum circuit, executed in two opposite directions. (If $E$ is
complex, the encoding and decoding circuits must also have opposite phase
shifts.)

The $2^n-1$ ``standard errors'' $|Z_0\ket\to|Z_a\ket$ are not the only
ones that can be corrected by the $E^\dagger$ decoding. Any error of
type

\beq |Z_0\ket\to U\,|Z_0\ket=\sum_a c_a\,|Z_a\ket, \eeq
is also corrected, since

\beq E^\dagger\,\sum_a c_a\,|Z_a\ket=\zz\otimes\sum_a c_a\,|a\ket, \eeq
is a direct product of \zz\ with the ancilla in some irrelevant
corrupted state. Note that {\it no knowledge of the syndrome is
needed\/} in order to correct the error [11]. Error correction is a
logical operation that can be performed automatically, without having
to execute quantum measurements. We know that the error is corrected,
even if we don't know the nature of that error.

It is essential that the result on the right hand side of (\theequation)
be a direct product. Only if the new ancilla state is the same for
$\zz=\0$ and $\zz=\1$, and therefore also for the complete computer
state in Eq.~(\ref{computer}), is it possible to coherently detach
the ancilla from the rest of the computer, and replace it by a fresh
ancilla (or restore it to its original state \ao\ by a dissipative
process involving still another, extraneous, physical
system).\footnote{The introduction of a dissipative process in the
quantum computer, which essentially is an analog device with a
continuous evolution, brings it a step closer to a conventional digital
computer!} This means, in the graphical formalism of quantum circuits,
that the ``wires'' corresponding to the old ancilla stop, and new
``wires'' enter into the circuit, with a standard quantum state for the
new ancilla.

There are many plausible scenarios for the emergence of coherent
superpositions of corrupted states, as in (\theequation). For example,
in an ion trap, a residual gas molecule, whose wave function is spread
over a domain much larger than the inter-ion spacing, can be scattered
by all the ions, as by a diffraction grating, and then all the ions are
left in a collective recoil state (namely, a coherent superposition of
states where one of the ions recoiled and the other ones did not).
Furthermore, {\it mixtures\/} of errors of type (\theequation) are also
corrigible. Indeed, if

\beq \rho=\sum_j p_j\,\sum_{ab}c_{ja}\,|Z_a\ket\,\bra Z_b|\,c^*_{jb},
\eeq
with $p_j>0$ and $\sum p_j=1$, then

\beq E^\dagger\rho\,E= \zz\,\bra z|\otimes
  \sum_j p_j\,\sum_{ab}c_{ja}\,|a\ket\,\bra b|\,c^*_{jb}, \eeq
again is a direct product of the logical qubit and the corrupted
ancilla.

These mixtures include the case where a physical qubit in the codeword 
gets entangled with an unknown environment, which is the typical
source of error. Let $\eta$ be the initial, unknown state of the
environment, and let its interaction with a physical qubit cause the
following unitary evolution:

\beq \begin{array}{lll}
 \0\otimes\eta & \to & \0\otimes\mu+\1\otimes\nu,\smallskip\\
 \1\otimes\eta & \to & \0\otimes\sigma+\1\otimes\tau,\end{array}
 \label{error} \eeq
where the new environment states $\mu,\ \nu,\ \sigma$, and $\tau$, are also
unknown, except for unitarity constraints. Now assume that the physical
qubit, that has become entangled with the environment in such a way, was
originally part of a codeword,

\beq |Z_0\ket=|X_{z0}\ket\otimes\0+|X_{z1}\ket\otimes\1. \eeq
That codeword, together with its environment, thus evolve as

\beq Z_0\otimes\eta\to Z'=X_{z0}\2\Bigl(\0\2\mu+\1\2\nu\Bigr)
  +X_{z1}\2\Bigl(\0\2\sigma+\1\2\tau\Bigr), \eeq
where I have omitted most of the ket signs, for
brevity. This can be written as\vspace{-2mm}

\beq \begin{array}{lll}
 Z' & = & 
  \Bigl[X_{z0}\2\0+X_{z1}\2\1\Bigr]\,{\displaystyle{\mu+\tau\over2}}+
  \Bigl[X_{z0}\2\0-X_{z1}\2\1\Bigr]\,{\displaystyle{\mu-\tau\over2}}\;+
  \smallskip \\ & &
  \Bigl[X_{z0}\2\1+X_{z1}\2\0\Bigr]\,{\displaystyle{\nu+\sigma\over2}}+
  \Bigl[X_{z0}\2\1-X_{z1}\2\0\Bigr]\,{\displaystyle{\nu-\sigma\over2}}\,.
  \end{array}\eeq
On the right hand side, the vectors

\beq \begin{array}{lll}
 Z_0 & = & X_{z0}\2\0+X_{z1}\2\1,\smallskip \\
 Z_r & = & X_{z0}\2\0-X_{z1}\2\1,\smallskip \\
 Z_s & = & X_{z0}\2\1+X_{z1}\2\0,\smallskip \\
 Z_t & = & X_{z0}\2\1-X_{z1}\2\0,\end{array} \label{Z}\eeq
correspond, respectively, to a correct codeword, to a phase error
($\1\to-\1$), a bit error ($\0\leftrightarrow\1$), which is the only
classical type of error, and to a combined phase and bit error.  If
these three types of errors can be corrected, we can also correct any
type of entanglement with the environment, as we shall soon see.

For this to be possible, it is necessary that the eight vectors in
Eq.~(\ref{Z}) be mutually orthogonal (recall that the index $z$ means 0
or 1).\footnote{There is a slight risk of confusion here, because the
same symbol 0 refers to the bit-value 0, and to the error free state of
a codeword. I see no way of circumventing this difficulty without
causing further confusion.} The simplest way of achieving this is to
construct the codewords $|0_0\ket$ and $|1_0\ket$ in such a way that
the following scalar products hold:

\beq \bra X_{zy}\,,\,X_{z'y'}\ket=
 \mbox{$1\over2$}\,\delta_{zz'}\,\delta_{yy'}. \label{XX}\eeq
(There are 10 such scalar products, since each index in this equation
may take the values 0 and 1.) If these conditions are satified, the
decoding of $Z'$ by $E^\dagger$ gives, by virtue of Eq.~(\ref{decod}),

\beq E^\dagger\,Z'=\zz\otimes\left(\ao\2{\mu+\tau\over2}
 +|r\ket\2{\mu-\tau\over2}+|s\ket\2{\nu+\sigma\over2}
 +|t\ket\2{\nu-\sigma\over2}\right).\eeq
The expression in parentheses is an entangled state of the ancilla and
the unknown environment. We cannot know it explicitly, but this is not
necessary: it is sufficient to know that it is the same state for
$\zz=\0$ or $\zz=\1$, or any linear combination thereof, as in
Eq.~(\ref{qubit}). We merely have to discard the old ancilla and bring
in a new one.

How to construct codewords that actually satisfy Eq.~(\ref{XX}), when
{\it any\/} one of their physical qubits is singled out, is a difficult
problem, best handled by a combination of classical codeword theory [1]
and finite group theory. I shall not enter into this subject here. I
only mention that in order to correct an arbitrary error in any one of
its qubits, a codeword must have at least five qubits: each one
contributes three distinct vectors, like $Z_r$, $Z_s$, and $Z_t$  in
Eq.~(\ref{Z}), and these, together with the error free vector $Z_0$,
make 16 vectors for each logical qubit value, and therefore $32=2^5$ in
the total. Longer codewords can correct more than one erroneous qubit.
For example, Steane's linear code [7], with 7 qubits, can correct not
only any error in a single physical qubit, but also a phase error,
$\1\to-\1$, in one of them, and a bit error, $\0\leftrightarrow\1$, in
another one (check!  $1+7\times3+7\times6=2^{7-1}$). A well designed
codeword is one where the orthogonal basis $|Z_a\ket$ corresponds to
the most plausible physical sources of errors.

The error correction method proposed above, in Eq.~(\ref{decod}), is
conceptually simple, but it has the disadvantage of leaving the logical
qubit \zz\ in a ``bare'' state, vulnerable to new errors that would be
not be detected. It is therefore necessary to re-encode that qubit
immediately, with another ancilla (or with the same ancilla, reset to
\ao\ by interaction with still another system). A more complicated but
safer method is to bring in a second ancilla, in a standard state
$|b=0\ket$, and have it interact with the complete codeword in such a
way that

\beq |Z_a\ket\otimes|b=0\ket\to |Z_0\ket\otimes|b=a\ket. \eeq
This is also a unitary transformation, which can be implemented by a
quantum circuit. Note that now the unitary matrix that performs that
error recovery is of order $2^{2n+1}$, instead of $2^{n+1}$.

Naturally, errors can also occur in the encoding and decoding process.
More sophisticated methods can however be designed, that allow fault
tolerant computation. An adaptive strategy is used, with several
alternative paths for error correction. Most paths fail, because new
errors are created; however, these errors can be detected, and there is
a high probability that one of the paths will eventually lead to the
correct result. As a consequence, the error correction circuits are
able to correct old errors faster than they introduce new ones. There
is then a high probability for keeping the number of errors small
enough, so that the correction machinery can successfully deal with
them [12].\\[7mm]

\noindent{\bf 4. Constrained dynamics}\bigskip

\noindent A quantum codeword is a redundant representation of a logical
qubit by means of several physical qubits. Since quantum codewords span
only a restricted subspace of the complete physical Hilbert space, the
unitary operations that generate quantum dynamics (that is, the
computational process) are subject to considerable arbitrariness. This
is most easily seen with the logical representation, $\zz\otimes\ao$. A
unitary transformation, $\one\otimes g$, where $g$ acts solely on the
ancilla's states, generates

\beq (\one\otimes g)\,\Bigl(|z\ket\otimes|a=0\ket\Bigr)=|z\ket\otimes
 \sum_a c_a\,|a\ket. \eeq
This is a corrupted, but corrigible codeword. In the physical
representation, this harmless unitary transformation becomes

\beq G=E\,(\one\otimes g)\,E^\dagger. \eeq
The unitary matrices $G$ are a representation (usually a reducible one)
of the U$n$ group. Consecutive applications of various transformations
of this type merely convert one corrigible error into another corrigible
error.  These transformations do not mix the two complementary
subspaces that represent the logical 0 and 1.

On the other hand, a genuine unitary transformation (one that is
actually needed for the computation) is, in the logical representation,
$\psi\to\psi'=(u\otimes\one)\psi$. It is encoded into

\beq U=E\,(u\otimes\one)\,E^\dagger, \eeq
for the physical representation.  Thus, in summary, all the ``legal''
unitary transformations are of type $E\,(u\otimes g)\,E^\dagger$, for
codewords that represent a single logical qubit.

For unitary transformations involving two logical qubits, the encoded
representation, including the possibility of corrigible errors, is
likewise

\beq U_{12}=(E_1\otimes E_2)\,[u_{12}\otimes(g_1\otimes g_2)]
  \,(E_1^\dagger\otimes E_2^\dagger), \eeq
where $u_{12}$ acts on the two logical qubits, and $g_1$ and $g_2$ act
on their respective ancillas. (I am assuming here that each logical
qubit is encoded separately, and that block coding is not used.) It is
obvious that in unitary transformations of that type, the logical steps
are not affected by the occurrence or evolution of corrigible errors.

Among these unitary transformations, there is a subgroup leaving
the zero-syndrome ancilla invariant (such a subgroup is called the {\it
little group\/} of the invariant state):

\beq g\,\ao=\ao.\eeq
Let us now focus our attention on these
transformations, that do not induce errors in correct codewords. They
only modify corrupted codewords, while keeping them corrigible. We may
imagine, if we wish, that error free codewords are stabilized by
erecting around them a high potential barrier: conceptually, we add to
the Hamiltonian a potential term, equal to zero for the legal codeword
states, and to a large positive number for erroneous states. This
artifice is similar to, but much simpler than, the use of the quantum
Zeno effect, that was proposed by several authors as a way of reducing
errors. It is actually not difficult to devise quantum circuits that
act like a potential barrier (the only serious difficulty is that such
a circuit must activate high frequency interactions with extraneous
qubits, and the latter may themselves be subject to errors, and induce
new ones).

In the logical basis, a ``legal'' (error free) state, $\zz\otimes\ao$,
which is invariant under the little group of \ao,
is recognized as being orthogonal to all $|z'\ket\otimes|a\neq0\ket$.
This can be written as an orthogonality relation

\beq \bra C_\alpha\,,\psi\ket=0, \eeq
where $C_\alpha$ is any linear combination of the various
$|z'\ket\otimes|a\ket$ with $a\neq0$. There are $2(2^n-1)$ linearly
independent $C_\alpha$, that span the ``illegal'' subspace (including
incorrigible errors). Let us normalize them by $\bra
C_\alpha\,,C_\beta\ket=\delta_{\alpha\beta}$.  After a legal unitary
evolution, $U\psi$ still is a legal state, and therefore

\beq \bra C_\alpha\,,U\psi\ket=0. \eeq
It follows that

\beq U\,C_\alpha=\sum_\beta A_{\alpha\beta}(U)\,C_\beta, \eeq
where the matrices $A_{\alpha\beta}(U)$ are a unitary representation of
$U$.  (If all legal $U$ are considered, that representation will not,
in general, be irreducible.)

It is also possible to construct Hermitian {\it operators\/} that
express the same constraints. Recall that the codewords are defined in
a Hilbert space with $2^{n+1}$ dimensions. Now consider

\beq M=\sum_{\alpha\beta}
 |C_\alpha\ket\,M_{\alpha\beta}\,\bra C_\beta|, \eeq
where $M_{\alpha\beta}$ is any matrix of order $2(2^n-1)$. Any legal
state obeys $M\psi=0$. Another constraint (for the same codeword) could
be $N\psi=0$, where

\beq N=\sum_{\alpha\beta} |C_\alpha\ket\,
 N_{\alpha\beta}\,\bra C_\beta|, \eeq
and $N_{\alpha\beta}$ is any other Hermitian matrix. It is easily shown
that

\beq [M,N]=iP, \eeq
where $P$ is still another Hermitian operator of the same type, and
satisfies $P\psi=0$ for all legal states.  Finally, we note that if
there are many logical qubits in the quantum computer, its state obeys
the nonlocal ``spacelike'' constraint equation

\beq M_1\otimes N_2\otimes\cdots\;\psi=0, \eeq
where the various operators refer to different codewords.

These equations are not completely trivial. They are like those
appearing in a quantum field theory with a gauge group. For example,
the canonical momenta of the free electro\-magnetic field are
$\pi^k=E^k$, where {\bf E} is the electric field vector. They satisfy
the constraint $\partial_k\pi^k=0$. This cannot hold as an operator
equation, because $\partial_k\pi^k$ does not commute with some other
field operators. However, a legal state vector (one without
``longitudinal photons'') obeys the constraint $\partial_k\pi^k\psi=0$.
The situation becomes more complicated for theories with non-Abelian
gauge groups, such as general relativity:  singular Schwinger terms
appear, and the factor ordering problem cannot be discussed without
regularization.\footnote{For a recent review, see ref.~[13].}

An important problem in quantum field theory (or, in general, in quantum
mechanics with constrained dynamical variables) is to properly define a
Hermitian scalar product. Should we include in it the spurious particles
that are generated by the gauge freedom, such as longitudinal photons?
When we consider codewords, the situation becomes simple and clear, as
we shall now see.

Consider indeed two different logical states of a quantum codeword, say

\beq {\mit\Phi}=E\,\Bigl(\phi\otimes\sum_a c_a\,|a\ket\Bigr), \eeq
and

\beq {\mit\Psi}=E\,\Bigl(\psi\otimes\sum_a c_a\,|a\ket\Bigr). \eeq
On the left hand side, there is the physical representation of the
codeword, and, in the parenthesis on the right hand side, its logical
representation. Note that, irrespective of the logical state ($\phi$ or
$\psi$), the ancilla has the same state $\sum c_a|a\ket$, because that
state represents the syndrome of the error, and the latter, caused by an
interaction with the environment, is independent of the logical state of
the qubit, as may be seen in Eq.~(\ref{Z}). It then readily follows from
the unitarity of $E$ that the scalar products,

\beq\bra{\mit\Phi},{\mit\Psi}\ket=\bra\phi,\psi\ket, \eeq
are the same for any two non-orthogonal states of a logical qubit, and
for their representation by codewords, even by corrupted ones. Further
work is in progress, in order to exploit the analogies of quantum
codeword dynamics with gauge field theory.\\[7mm]

\noindent{\bf Acknowledgments}\bigskip

\noindent I am grateful to Peter Shor and Andrew Steane for clarifying
remarks. This research was supported in part by the National Science
Foundation under Grant No.\ PHY94-07194.

\bigskip\noindent{\bf References}\frenchspacing\small

\begin{enumerate}
\setlength{\leftmargin}{-.25in}
\item D. Welsh, {\it Codes and Cryptography\/}, Oxford University Press
(1989), Chapt. 4.
\item A. Peres, {\it Quantum Theory: Concepts and Methods\/}, Kluwer,
Dordrecht (1993), Chapt. 5.
\item J. I. Cirac and P. Zoller, Phys. Rev. Lett. 74 (1995) 4091.
\item W. K. Wootters and W. H. Zurek, Nature 299 (1982) 802.
\item P. W. Shor, Phys. Rev. A 52 (1995) 2493.
\item R. Laflamme, C. Miquel, J. P. Paz, and W. H. Zurek, Phys. Rev.
Lett. 77 (1996) 198.
\item A. M. Steane, Phys. Rev. Lett. 77 (1996) 793; Proc. Roy. Soc.
(London) in press.
\item C. H. Bennett, D. P. DiVincenzo, J. A. Smolin, and W. K. Wootters,
Phys. Rev. A (in press).
\item E. Knill and R. Laflamme, ``A theory of quantum error-correcting
codes'' (Los Alamos report LA-UR-96-1300).
\item H. Goldstein, {\it Classical Mechanics\/}, Addison-Wesley, Reading
(1980), Chapt. 6.
\item A. Peres, Phys. Rev. A 32 (1985) 3266.
\item P. W. Shor, ``Fault tolerant quantum computation'' in {\it Proc.
37th Symposium on Foundations of Computer Science\/} (1996) in press.
830.
\item N. C. Tsamis and R. P. Woodard, Phys. Rev. D 36 (1987) 3641.

\end{enumerate} \end{document}